\documentclass[a4paper,11pt]{article}
\pdfoutput=1 

\usepackage{jcappub} 
\usepackage{soul}

\usepackage[T1]{fontenc} 
\usepackage[dvipsnames]{xcolor}
\usepackage{color}

\title{Dark sector interactions in the $w \rightarrow -1$ limit: velocity locking in pure momentum exchange models}

\author[a]{Nathan Cruickshank,}
\author[a]{Robert Crittenden,}
\author[a, b, c]{Kazuya Koyama,}
\author[a]{and Marco Bruni}

\affiliation[a]{Institute of Cosmology and Gravitation, University of Portsmouth,\\
Dennis Sciama Building, Burnaby Road, Portsmouth, PO1 3FX, United Kingdom}
\affiliation[b]{Kavli IPMU (WPI), UTIAS, The University of Tokyo, Kashiwa, Chiba 277-8583, Japan}
\affiliation[c]{Yukawa Institute for Theoretical Physics, Kyoto University, Kyoto 606-8502, Japan}

\emailAdd{nathan.cruickshank@port.ac.uk}
\emailAdd{robert.crittenden@port.ac.uk}
\emailAdd{kazuya.koyama@port.ac.uk}
\emailAdd{marco.bruni@port.ac.uk}

\abstract{Models of interacting dark energy (DE) and dark matter (DM) involving pure momentum exchange are a promising avenue for resolving cosmological tensions. However, the behaviour of these interactions in the theoretically challenging limit where the DE equation of state, $w$, approaches $-1$ is not fully understood. We demonstrate that a generic feature of these models is a $w$-dependent velocity-locking mechanism, which systematically shifts the onset of matter power spectrum suppression to smaller scales as $w \rightarrow -1$. The suppression magnitude depends on the difference in fluid velocities. In this limit, however, the interaction's drag dominates over the DE pressure support and causes the DE velocity to track that of the DM fluid at larger scales. This mechanism provides a physical explanation for the weaker constraints found in the literature when $w\approx-1$ in models where the interaction strength does not explicitly depend on $w$. We also demonstrate that the common approximation of neglecting DE perturbations ($\delta_{\mathrm{DE}}=\theta_{\mathrm{DE}}=0$) fails in this limit. By artificially increasing the velocity difference between the fluids, this simplification incorrectly removes the $w$-dependent velocity-locking mechanism and erases the shift in power spectrum suppression to smaller scales. This leads to an overestimation of the constraining power of cosmological data on the interaction strength.}

\begin{document}
\maketitle
\flushbottom

\section{Introduction}
\label{sec:intro}

The standard cosmological model, $\Lambda$CDM, has been successful in describing a wide range of cosmological observations, from the cosmic microwave background (CMB) to the large-scale structure of the universe. However, as the precision of cosmological data has improved, tensions have emerged between different observational probes. Most notable are the Hubble tension, a discrepancy in the measured expansion rate of the universe today \cite{hu2023hubble,Abdalla:2022yfr,Riess_2016,Riess_2022,Stiskalek:2025ktq}, and to a lesser extent, the $S_8$ tension, which relates to the amplitude of matter clustering \cite{Planck_2018,Hildebrandt_2020,Asgari_2021,Di_Valentino_2021,Abbott_2022}. Recent results from the late-time KiDS survey are consistent with those of the Planck CMB experiment \cite{Wright:2025xka,Stolzner:2025htz}; however, other surveys consistently infer a value of $S_8$ that is $2$--$3\sigma$ lower than that expected from Planck measurements \cite{CosmoVerseNetwork:2025alb}. These discrepancies suggest that our understanding of the late evolution of the universe and its explanation through dark components in the framework of General Relativity remains incomplete, pointing to the possible need for new physics.

To help resolve these apparent tensions, interacting dark energy (DE) and dark matter (DM) models have been proposed, where the components of the dark sector are not entirely independent but instead interact with each other in ways other than gravitational \cite{Salvatelli:2014zta,Bolotin_2015,Wang_2016,Abdalla:2022yfr,Tamanini_2015,Pourtsidou:2016ico,Di_Valentino_2017,Pan_2019,Pan_2020_1,Pan_2020_2,Amendola_2020,Di_Valentino_2020_1,Di_Valentino_2020_2,Khoury:2025txd}. 
DM-DE interactions can be broadly categorised into models with energy transfer, which modify the background evolution of the universe, and those with only pure momentum exchange, which primarily affect the growth of cosmic structures. Pure momentum exchange models offer a compelling way to address the $S_8$ tension. 

A pure momentum exchange is analogous to the non-relativistic limit of a more general dark sector interaction, much like how Thomson scattering is the non-relativistic limit of Compton scattering. Momentum exchange can be defined as long as $w \neq -1$, where $w=p/\rho$ is the DE equation of state. When $w = -1$ exactly, the fluid behaves as vacuum energy. If this vacuum energy does not interact, it is a true cosmological constant; if it does, however, its energy density must evolve \cite{Wands:2012vg,Martinelli:2019dau,Hogg:2020rdp,Sebastianutti:2023dbt}. In this limit, the DE Euler equation breaks down because the fluid lacks a well-defined four-velocity and cannot support a velocity divergence. Instead, when $w = -1$ the interaction must be described through a constraint equation derived from energy-momentum conservation. As shown in \cite{Wands:2012vg}, the momentum transfer in this case satisfies
\begin{equation}
\label{eq:constraint}
    -f = \delta V + \dot{V} \theta_{\mathrm{DM}}\:,
\end{equation}
where $f$ is the momentum transfer, $V$ is the background vacuum energy density, $\delta V$ is its perturbation, and $\theta_{\mathrm{DM}}$ is the velocity divergence of the dark matter component. This constraint shows that a pure momentum exchange cannot be sustained at $w = -1$ unless $\dot{V} \neq 0$, which would imply an accompanying energy transfer in the background.

Current DESI measurements indicate that the effective dark energy equation of state $w(z)$ may have transitioned between the phantom ($w<-1$) and non-phantom ($w>-1$) regimes during its evolution across cosmic time \cite{DESI:2025zgx}. Such a ``phantom crossing'' cannot be explained by a non-interacting single canonical scalar field or perfect fluid, suggesting an interaction in the dark sector. In particular, it is important to explore and understand the effects of dark scattering interactions when $w \rightarrow -1$. This limit has previously been shown to be theoretically challenging, as many interacting fluid models suffer from large-scale instabilities that require specific conditions of the coupling to be avoided \cite{Valiviita:2008iv,Jackson:2009mz}. In this work, we explore the fluid dynamics of this regime within a pure momentum exchange framework that is free from these large-scale instabilities.

\begin{figure}[tbp]
\centering
\includegraphics[width=0.45 \textwidth]{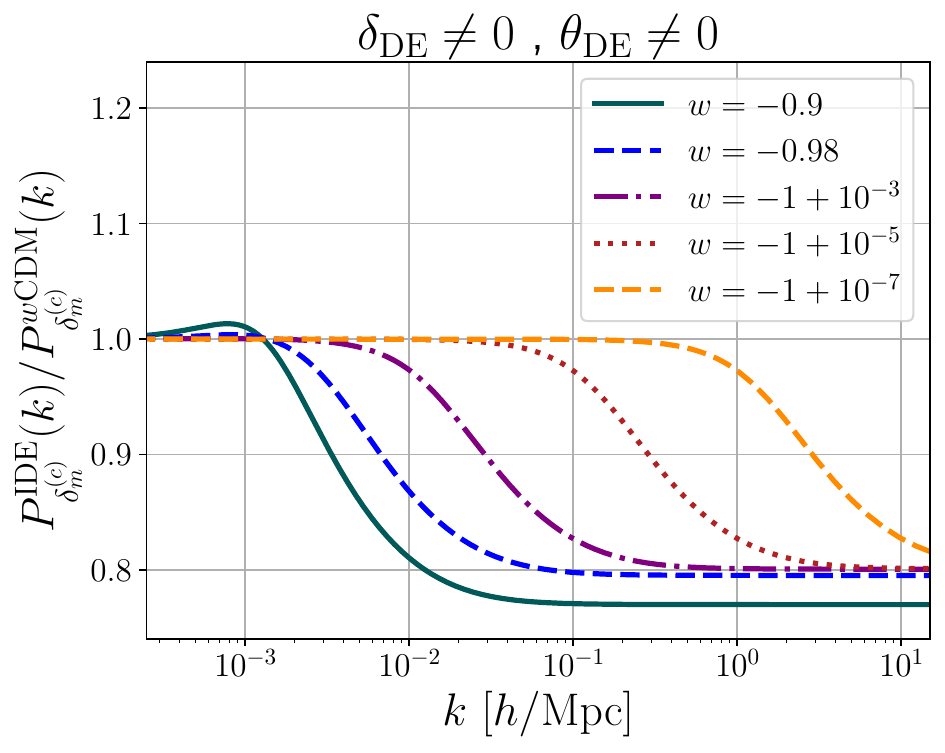}
\includegraphics[width=0.45 \textwidth]{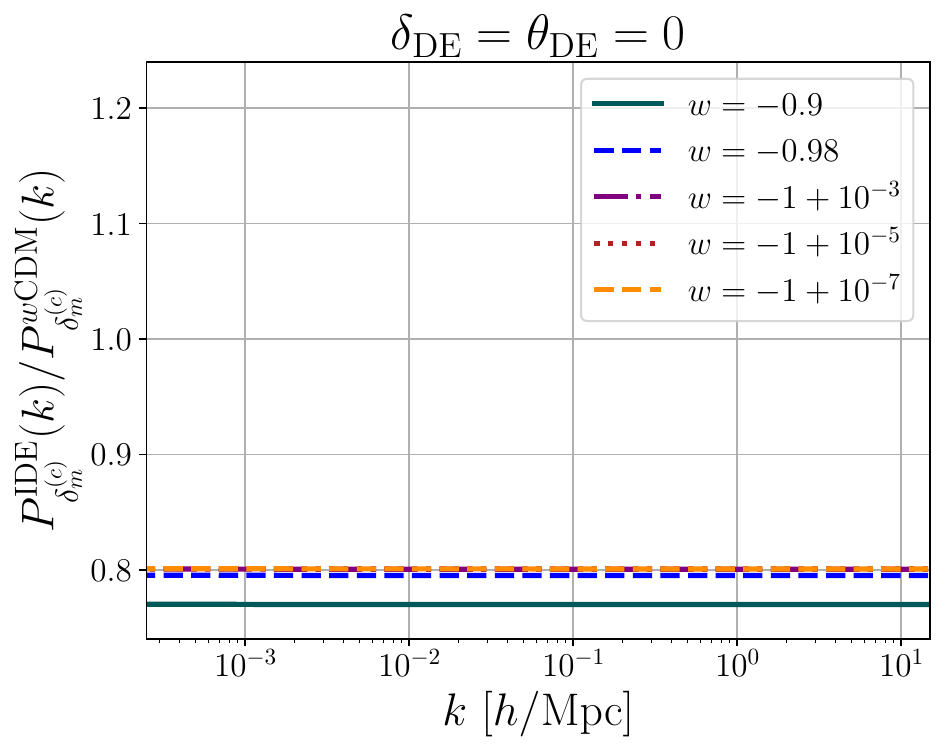}
\hfill
\caption{\label{fig:pk_theta_de_off} 
Ratio of the linear matter power spectrum (in the comoving-synchronous gauge) with interaction to that of a non-interacting $w$CDM model, computed for a variety of $w$ values at $z = 0$ using the dark scattering model with an interaction strength of $A = 15$ [b/GeV]. \textbf{Left:} interaction and $w$CDM models where $\delta_{\mathrm{DE}} \neq 0$ and $\theta_{\mathrm{DE}} \neq 0$. \textbf{Right:} interaction and $w$CDM models where $\delta_{\mathrm{DE}} = \theta_{\mathrm{DE}} = 0$ (see Section \ref{sec:no_de}).}
\end{figure}

In particular, we will focus on the matter power spectrum in such models, as is illustrated in Figure \ref{fig:pk_theta_de_off}. As can be seen from the left-hand panel, there is a scale-dependent suppression of the power on smaller scales, which results from a suppression of the DM velocity due to its interaction with DE. We will demonstrate that the onset scale of this suppression systematically shifts to smaller physical scales (higher $k$) as $w$ approaches $-1$, and that this behaviour is a generic feature of momentum-exchange models. We will also investigate the impact that ignoring DE velocity can have on parameter constraints when $w\approx-1$.

We will then explore the underlying physics of this effect.
We show that the interaction depends on the difference between the DM and DE velocities, driving them towards each other. While the DE velocity is small, the interaction tends to suppress the DM velocity and therefore its perturbations on small scales. However, when the interaction is sufficiently strong, the DE velocity is drawn towards the DM, effectively disabling the suppression mechanism.  

\subsection{Gauge choice}
Throughout this paper, we work in the conformal Newtonian gauge, the line element of which is 
\begin{equation}
    ds^2 = a^2(\tau)\left\{-(1+2\psi)d\tau^2+(1-2\phi)dx^idx_i\right\}\:,
\end{equation}
where $\tau$ is conformal time. However, we choose to plot the power spectrum in terms of the comoving-synchronous gauge of the total matter density, $\delta^{(c)}_m$, as the Newtonian gauge shows a strong enhancement at small $k$ \cite{Dent:2008ia,Bruni:2011ta}. We do this by plotting the power spectrum of the gravitational potential, $\psi$. In the Newtonian gauge, $\psi$ directly relates to the matter density contrast through
\begin{equation}
    \delta_m = -2\left[1+\frac{k^2}{3(aH)^2}\right]\psi\:,
\end{equation}
where $a$ is the scale factor, $H$ is the Hubble parameter and $k$ is the comoving wavenumber \cite{Ma_1995}. Comparing this to the same relation in the comoving-synchronous gauge, 
\begin{equation}
    \delta^{(c)}_m = -\frac{2k^2\psi}{3(aH)^2}\:,
\end{equation}
shows significant gauge-dependent differences on larger scales ($k\lesssim aH$). It also demonstrates that $\psi$ sources the comoving density field $\delta^{(c)}_m$ as it does in Newtonian gravity on all scales.

\section{Pure momentum exchange models}
\label{sec:models}

Although some interaction models treat DE as a scalar field \cite{Pourtsidou_2013,Skordis_2015,Pourtsidou:2016ico,palma2023cosmological}, this work focuses on the approach in which both DE and DM are described as fluids. However, we restrict our analysis to the non-phantom regime ($w \ge -1$) because the chosen fluid model phenomenologically mimics a single canonical scalar field, for which a phantom equation of state would imply unphysical negative kinetic energy. In the conformal Newtonian gauge, the dark matter--dark energy scattering is generically described by the following continuity and Euler equations \cite{Simpson:2010vh,Asghari:2019qld}, where interaction terms appear on the right-hand side of the latter:
\begin{equation}
\label{eq:delta_dm}
    \delta'_{\mathrm{DM}}
    +\theta_{\mathrm{DM}}
    -3\phi' = 0\:,
\end{equation}
\begin{equation}
\label{eq:theta_dm}
    \theta'_{\mathrm{DM}}
    +\mathcal{H}\theta_{\mathrm{DM}}
    -k^2\psi = 
    -\Gamma(a)\left(\theta_{\mathrm{DM}}-\theta_{\mathrm{DE}}\right)\:,
\end{equation}
\begin{equation}
\label{eq:delta_de}
    \delta'_{\mathrm{DE}}
    +\left[(1+w)+9(1+w)(c_s^2-c_a^2)\frac{\mathcal{H}^2}{k^2}\right]\theta_{\mathrm{DE}} 
    +3(c_s^2-w)\mathcal{H}\delta_{\mathrm{DE}}
    -3(1+w)\phi' = 0\:,
\end{equation}
\begin{equation}
\label{eq:theta_de}
    \theta'_{\mathrm{DE}}
    +\left(1-3c_s^2\right)\mathcal{H}\theta_{\mathrm{DE}}
    -\frac{c_s^2 k^2}{\left(1+w\right)}\delta_{\mathrm{DE}}
    -k^2\psi = 
    \Gamma(a)R(a)\left(\theta_{\mathrm{DM}}-\theta_{\mathrm{DE}}\right)\:.
\end{equation}
Here, a prime denotes a derivative with respect to conformal time, $\delta_i$ is the density contrast for fluid $i$ (DM or DE), $\theta_i$ is its velocity divergence, and $\rho_i$ is its background energy density. $\phi$ and $\psi$ are the gravitational potentials, and the terms proportional to the conformal Hubble parameter, $\mathcal{H} = a'/a$, represent Hubble friction. The DE component is described by its equation of state $w$, its physical sound speed squared $c_s^2$, and its adiabatic sound speed squared $c_a^2$. For a constant $w$, the adiabatic sound speed is simply $c_a^2=w$. We use the cosmological units of $\textrm{Mpc}^{-2}$ for densities and $\textrm{Mpc}$ for distances.

The right-hand sides of the Euler equations (Equations \ref{eq:theta_dm} and \ref{eq:theta_de}) model the momentum transfer between the dark fluids. The interaction depends on the difference in the velocity divergence of the two fluids $(\theta_{\mathrm{DM}} - \theta_{\mathrm{DE}})$, and vanishes when the velocities match. It was shown in \cite{Asghari:2019qld} that the growth of DM density perturbations is directly tied to this velocity difference, with $\delta^\prime_{\mathrm{DM}}\simeq\theta_{\mathrm{DE}}-\theta_{\mathrm{DM}}$.

The interaction rate $\Gamma(a)$, which has dimensions of inverse distance (or time) and units of $\textrm{Mpc}^{-1}$, is generically time-dependent and depends on the specific model being considered (see below). The momentum ratio $R(a)$ is given by
\begin{equation}
 R(a) = \frac{\rho_{\mathrm{DM}}(a)}{\left(1+w\right)\rho_{\mathrm{DE}}(a)}\:,
\end{equation}
and arises from momentum conservation. This momentum exchange manifests as a drag force on the fluid perturbations. For an interaction where $\Gamma(a) > 0$, this leads to a suppression in the growth of DM density perturbations and thus a suppression of the large-scale structure growth rate, which helps address the $S_8$ tension.

\subsection{The dark scattering and late-time interaction models}

Here, we analyse two distinct parametrisations from the literature to confirm that the observed $w$-dependent suppression scale shift in the power spectrum is a robust prediction for this class of interaction.
The first is the dark scattering model, which describes the interaction as analogous to Thomson scattering. It was proposed in \cite{Simpson:2010vh} and further explored in \cite{Baldi:2016zom,Kumar:2017bpv,Bose:2017jjx,Asghari_2020,BeltranJimenez:2021wbq,Carrilho:2021hly,Linton_2022,Poulin:2022sgp,Carrilho:2022mon,Lague:2024sox,Carrion:2024jur,Tsedrik:2022cri,Cruickshank:2025iig}. The second is a model we refer to here as the late-time interaction model, described in \cite{Asghari:2019qld} and subsequently explored in \cite{BeltranJimenez:2021wbq,Figueruelo:2021elm,BeltranJimenez:2022irm,Poulin:2022sgp,BeltranJimenez:2024lml,Lague:2024sox,ACT:2025tim}.

In the dark scattering (DS) model, the interaction rate is given by 
\begin{equation} 
\label{eq:gamma_ds}
\Gamma_{\mathrm{DS}} (a) \equiv a {\rho_{\mathrm{DE}}(a) \xi(1+w)}, 
\end{equation}
where $\xi = \sigma_\mathrm{D}/m_{\mathrm{DM}}$ is the ratio of the interaction cross-section $\sigma_\mathrm{D}$ to the DM particle mass $m_{\mathrm{DM}}$ (in units of barns/GeV) when we set $c=1$. 
In practice, we reparametrise this using $A \equiv \xi(1+w)$, which we convert to units of $\textrm{Mpc}$ in our code implementation. This reparametrisation is crucial for analysing the model near $w = -1$. If $\xi$ were used as the primary model parameter, the full interaction term, proportional to $\xi(1+w)$, would vanish as $w \to -1$ for a fixed $\xi$. This would render $\xi$ observationally unconstrained in the cosmological constant limit. Using $A$ as the fundamental parameter ensures the model predicts a well-defined and measurable interaction strength as this limit is approached. Maintaining a finite $A$ in this limit necessarily requires $\xi$ to become arbitrarily large, leading to a potential breakdown of the underlying microphysics. However, we adopt this parametrisation to maintain consistency with the existing literature and to fully explore the theoretical issues that arise in this regime.

\begin{figure}[tbp]
\centering
\includegraphics[width=0.7 \textwidth]{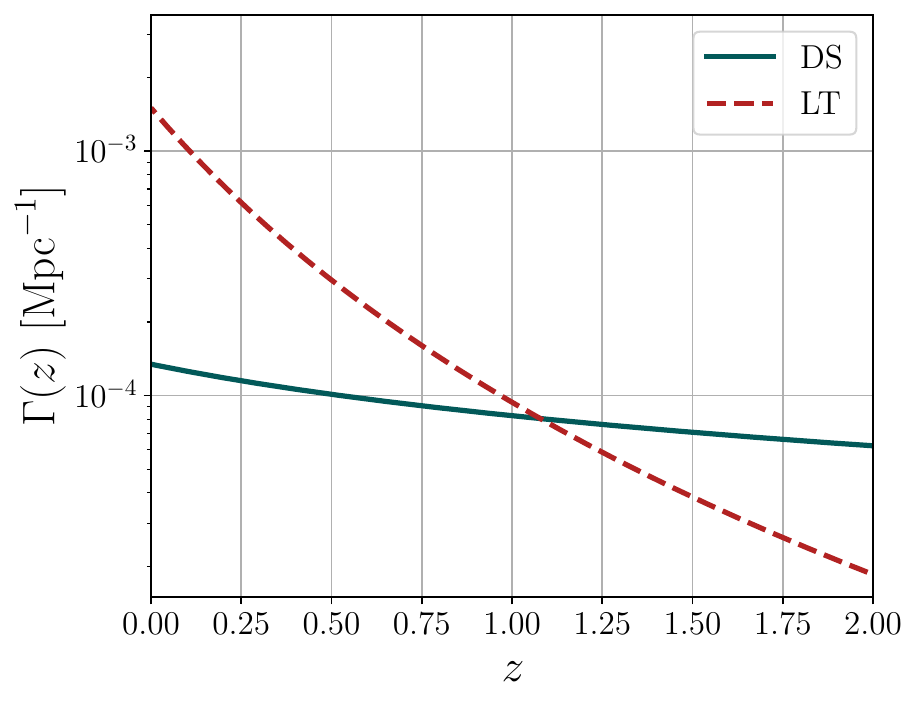}
\hfill
\caption{\label{fig:Gamma_a} 
A comparison of the $\Gamma(z)$ evolution for the dark scattering (DS) and late-time (LT) interaction models when $c=1$, $A = 15$ [b/GeV] and $\Gamma_{\mathrm{LT}}=2{H_0}^3$.}
\end{figure}

In the alternative late-time (LT) interaction approach, the interaction rate is given by
\begin{equation}
\label{eq:gamma_lt}
    \Gamma_{\mathrm{LT}}(a) \equiv \frac{a\Gamma_{\mathrm{LT}}}{\rho_{\mathrm{DM}}(a)} \:,
\end{equation}
and is intentionally designed to strengthen at low redshifts. This parametrisation, first introduced in \cite{Asghari:2019qld}, expresses the interaction rate in terms of background densities and $a$. The time-dependent interaction rate for this model, $\Gamma_{\mathrm{LT}}(a)$, is constructed from a constant physical parameter, $\Gamma_{\mathrm{LT}}$, which has units of $\mathrm{Mpc}^{-3}$. A comparison of the evolution of $\Gamma(a)$ for both the dark scattering and late-time interaction models can be seen in Figure \ref{fig:Gamma_a}.

The dark scattering model's dependence on $\rho_{\mathrm{DE}}(a)$ causes the interaction to remain approximately constant in time, whereas the dependence of the LT model on $1/\rho_{\mathrm{DM}}(a)$ results in a coupling that strengthens very quickly at late times. Alternatively, one can parametrise the interaction to allow for a general time-dependence. We demonstrate this in \cite{Cruickshank:2025iig} by treating the coupling strength as a free function of time and exploring how well it can be constrained in different redshift bins.

A common approximation made in the literature, particularly for simulation-based approaches, is to treat DE fluctuations as negligible ($\delta_{\mathrm{DE}} = \theta_{\mathrm{DE}}=0$). While not physically consistent, given the assumed sound speed of $c_s=1$ \cite{Baldi:2014ica,Baldi:2016zom,Bose:2017jjx,Carrilho:2021hly,Tsedrik:2022cri,Carrilho:2022mon,BeltranJimenez:2022irm}, the DE fluctuations are very suppressed, making this a reasonable approximation in many contexts. However, we will show below that this approximation leads to incorrect results in the limit that $w \rightarrow -1$.

\subsection{Scale-dependent power spectra}

To study the impact of momentum exchange in the dark scattering model, we have implemented the model's modified Euler equations into the Einstein-Boltzmann solver code \texttt{CLASS} \cite{Blas_2011}. Initially, we fix the DE sound speed to $c_s^2=1$, and choose a large coupling strength of $A = 15$ [b/GeV] to clearly illustrate the physical effects. We initiate the interaction at redshift $z = 10$; we tested this starting redshift and confirmed it has a minimal impact on our results as the interaction's effect on structure growth is subdominant until later times when the DE density becomes significant. To test the $w \to -1$ limit, we use a constant equation of state $w=-1+10^{-5}$, which we compare against a less extreme case of $w=-0.9$. This allows us to isolate the effect of $w$ on the fluid dynamics and the linear matter power spectrum.

\begin{figure}[tbp]
\centering
\includegraphics[width=0.45 \textwidth]{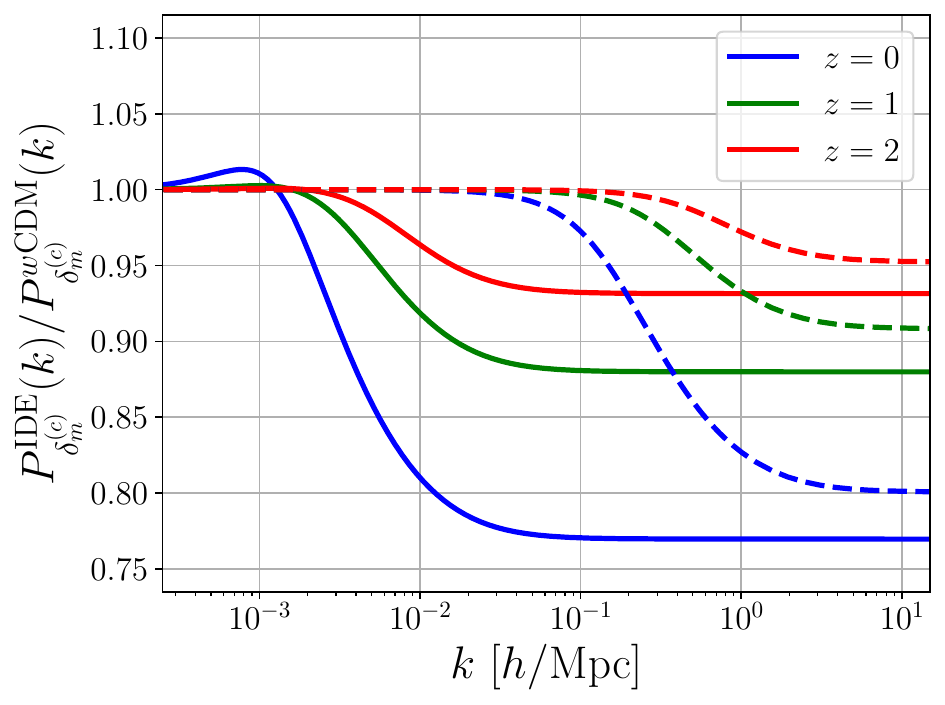}
\includegraphics[width=0.45 \textwidth]{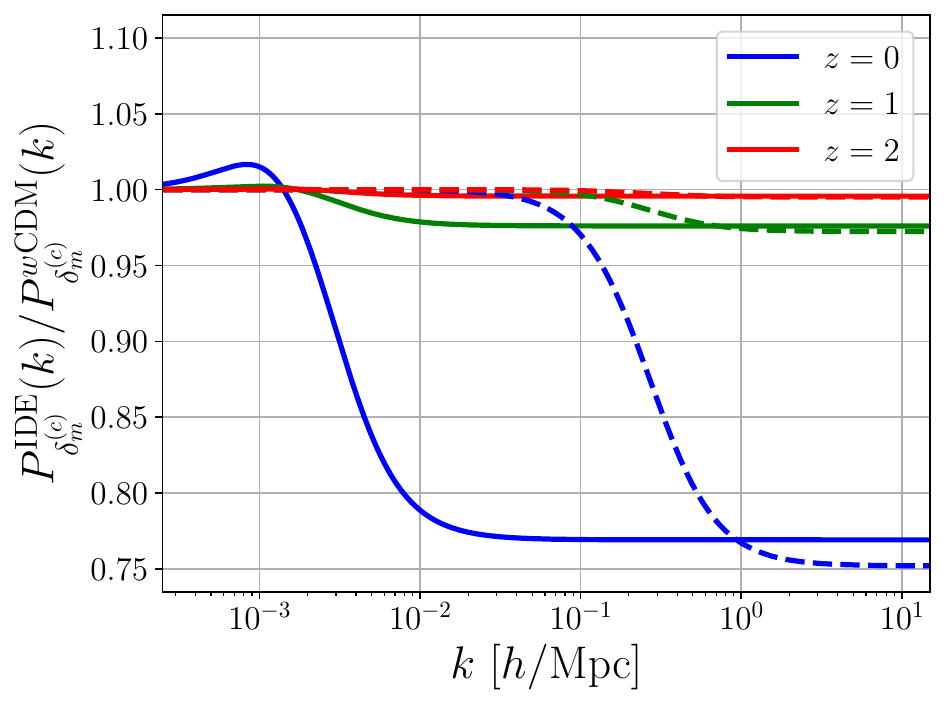}
\hfill
\caption{\label{fig:pk}
Ratio of the linear matter power spectrum in the comoving-synchronous gauge with an interaction to that of a non-interacting $w$CDM model, shown at redshifts $z = 0, 1, 2$. Results are shown for $w=-0.9$ (solid) and $w=-1+10^{-5}$ (dashed). \textbf{Left:} The dark scattering model when $A = 15$ [b/GeV]. \textbf{Right:} The late-time interaction model {when $\Gamma_{\mathrm{LT}} = 2{H_0}^3$}.}
\end{figure}

Figure \ref{fig:pk} (left panel) presents the ratio of the linear matter power spectrum in the dark scattering model to that of a non-interacting $w$CDM cosmology. The ratio is shown for $w=-0.9$ (solid lines) and $w=-1+10^{-5}$ (dashed lines) at three different redshifts. The overall suppression of power is reduced in the $w=-1+10^{-5}$ case, as the interaction strength is proportional to $\rho_{\mathrm{DE}}(a)$. Since $\rho_{\mathrm{DE}} \propto a^{-3(1+w)}$, a value of $w \approx -1$ leads to a nearly constant DE density, reducing the overall interaction strength compared to the $w=-0.9$ case where $\rho_{\mathrm{DE}}$ grows more rapidly into the past.

On large scales (low $k$), the interaction becomes ineffective for both values of $w$. This scale dependence arises because the peculiar velocities of the DM and DE fluids coincide on super-horizon scales, causing the velocity divergence difference ($\theta_{\mathrm{DM}}-\theta_{\mathrm{DE}}$) to vanish, as noted in \cite{Asghari:2019qld,BeltranJimenez:2020qdu,Chamings:2019kcl}. However, we find that the scale at which this suppression begins is not fixed, but instead shifts to higher $k$ as $w \to -1$.

Similarly, for the late-time interaction model, we use a modified version of \texttt{CLASS} as described in \cite{Lague:2024sox}. To study its distinct fluid dynamics, we again set the sound speed to $c_s^2 = 1$ and choose a value of $\Gamma_{\mathrm{LT}} = 2{H_0}^3$ for the constant coupling parameter. This value was chosen to closely match the suppression magnitude of the dark scattering model. The results for this model are shown in the right panel of Figure \ref{fig:pk}. While it exhibits a similar overall suppression of power to the dark scattering model, its redshift evolution is notably different. The interaction has a much weaker effect at high redshift. This is an expected consequence of the model's structure, as its coupling strength grows at late times, in direct contrast to the dark scattering model. While the overall suppression of power is reduced in the $w=-1+10^{-5}$ case for the dark scattering model, the opposite is true for this model. This is due to $\rho_{\mathrm{DE}}(a)$ appearing in the denominator of the interaction term when substituting Equation \ref{eq:gamma_lt} into Equation \ref{eq:theta_de}.

\begin{figure}[tbp]
\centering
\includegraphics[width=0.45 \textwidth]{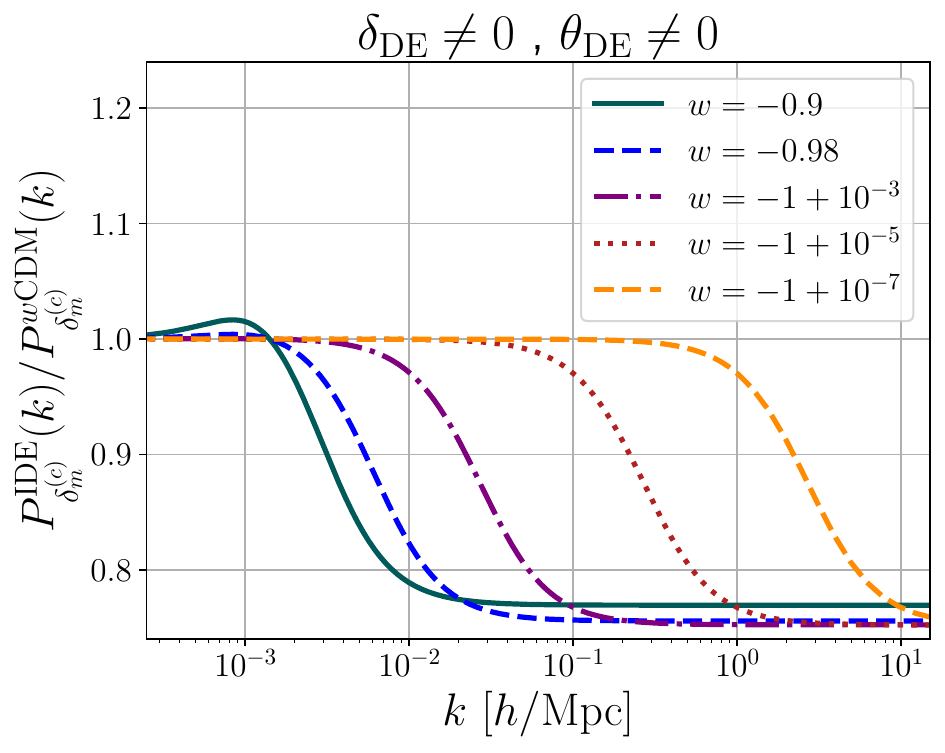}
\includegraphics[width=0.45 \textwidth]{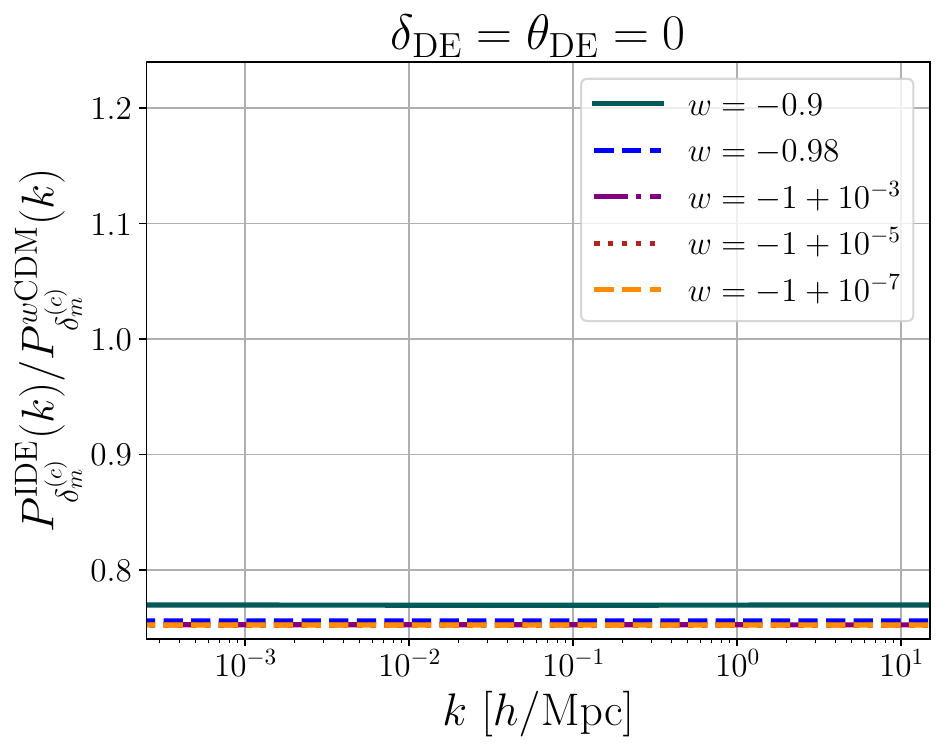}
\hfill
\caption{\label{fig:pk_theta_de_off_gamma} 
Ratio of the linear matter power spectrum in the comoving-synchronous gauge with interaction to that of a non-interacting $w$CDM model, computed for a variety of $w$ values at $z = 0$ using the {late-time interaction} model when $\Gamma_{\mathrm{LT}} = 2{H_0}^3$. \textbf{Left:} interaction and $w$CDM models where $\delta_{\mathrm{DE}}\ne0$ and $\theta_{\mathrm{DE}}\ne0$. \textbf{Right:} interaction and $w$CDM models where $\delta_{\mathrm{DE}} = \theta_{\mathrm{DE}} = 0$ (see Section \ref{sec:no_de}).}
\end{figure}

Notably, as in the dark scattering case, the scale at which the power spectrum suppression begins in Figure \ref{fig:pk} is shifted to higher $k$ in the $w=-1+10^{-5}$ case compared to the $w=-0.9$ case. This is further demonstrated in Figure \ref{fig:pk_theta_de_off_gamma}. As in Figure \ref{fig:pk_theta_de_off}, the left panel shows that when DE perturbations are included ($\delta_{\mathrm{DE}}\ne0$ and $\theta_{\mathrm{DE}}\ne0$), the scale at which the power spectrum suppression begins is shifted to higher $k$ as $w \to -1$. In contrast, the right panel shows that when these perturbations are set to zero ($\delta_{\mathrm{DE}} = \theta_{\mathrm{DE}} = 0$), this $w$-dependence vanishes and the suppression becomes scale-independent.

\subsection{Effect on growth rate}

The $w$-dependent nature of the suppression scale has a direct impact on the growth of large-scale structure, which can be constrained by measurements of redshift-space distortions (RSD). A key observable derived from RSD is the combination $f\sigma_8$, which measures the rate of structure growth. The logarithmic growth rate is defined as $f(a) \equiv \frac{d \ln D}{d \ln a}$, where $D(a)=\frac{\delta(a)}{\delta(a=1)}$ is the growth factor. This is combined with $\sigma_8$, which describes the amplitude of matter perturbations in an $8h^{-1}\,\mathrm{Mpc}$ radius sphere.
Due to its sensitivity to the dynamics of DM perturbations, $f\sigma_8$ is an excellent probe for the effects of an evolving dark sector interaction. Recent forecasts have demonstrated its potential to place tight constraints on the time-dependent coupling strength of momentum exchange models \cite{Cruickshank:2025iig}. Such measurements could also be used to constrain the strength of an interaction when the value of $w$ evolves. 

\begin{figure}[tbp]
\centering
\includegraphics[width=0.45 \textwidth]{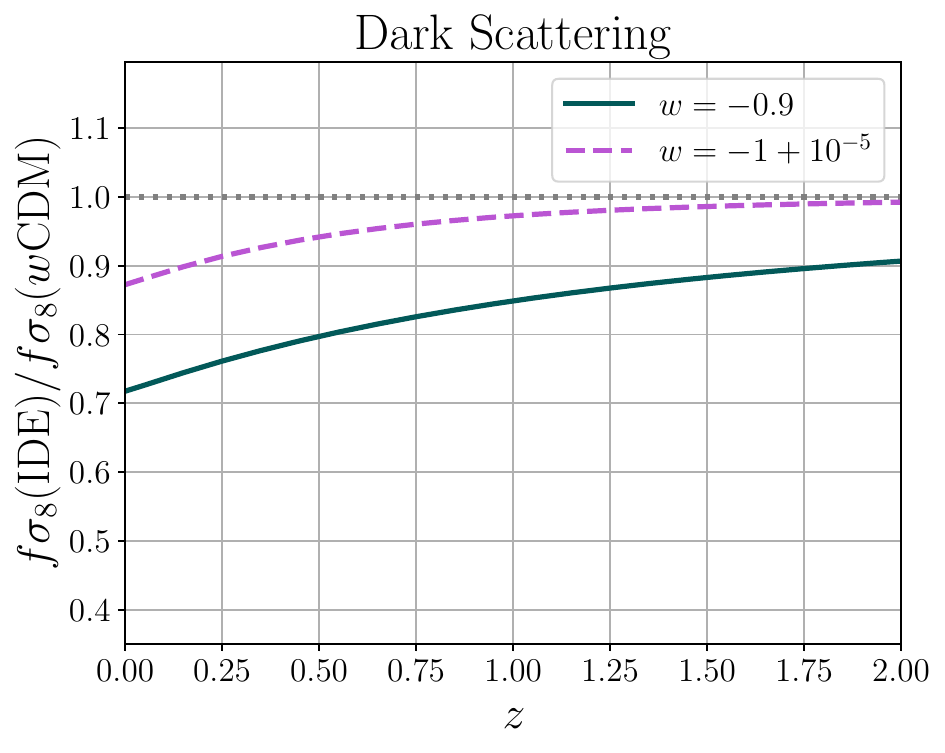}
\includegraphics[width=0.45 \textwidth]{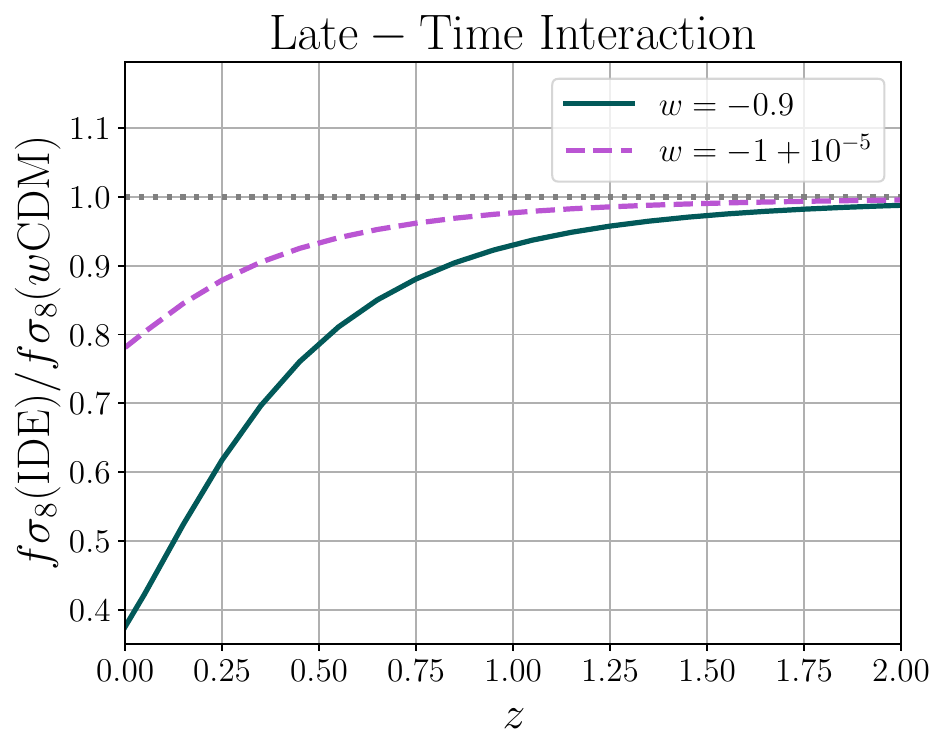}
\hfill
\caption{\label{fig:scale_indep_fsigma8}
Ratio of $f\sigma_8$ in the interacting model to that in the fiducial $w$CDM case, shown as a function of redshift. Results are shown for $w=-0.9$ (solid) and $w=-1+10^{-5}$ (dashed). \textbf{Left:} The dark scattering model when $A = 15$ [b/GeV]. \textbf{Right:} The {late-time interaction} model {when $\Gamma_{\mathrm{LT}} = 2{H_0}^3$}.}
\end{figure}

The suppression of the linear matter power spectrum, shown in Figure \ref{fig:pk}, directly translates to a suppression of the growth rate of structure. The ratio of $f\sigma_8$ in the interacting models to the non-interacting case is shown in Figure \ref{fig:scale_indep_fsigma8} for both the dark scattering and {late-time interaction} models. In both models, the suppression of $f\sigma_8$ is significantly weaker when $w \approx -1$. This is a direct consequence of the scale dependence. Since the interaction affects only the smaller scales, its overall impact on $\sigma_8$ and therefore the growth rate is reduced.

\section{The dynamics of the \texorpdfstring{\boldmath{$w\to-1$}}{w -> -1} limit}
\label{sec:dynamics}

Having established that the systematic shift in the suppression scale is a generic feature of momentum exchange models, we now deconstruct the underlying physical mechanism; for simplicity, here we focus on the dark scattering parametrisation. A natural expectation might be that the strength of the interaction is independent of $w$ when the DM Euler equation (Equation \ref{eq:theta_dm}) lacks an explicit $(1+w)$ dependence. However, as shown in the left-hand panels of Figures \ref{fig:pk_theta_de_off} and \ref{fig:pk_theta_de_off_gamma}, the overall suppressive effect clearly weakens as the DE equation of state approaches $-1$ for both models. The origin of this $w$-dependent behaviour in both models lies in the dynamics of the DE fluid itself. This leads to a velocity-locking phenomenon, similar to that in the strong coupling regime, that becomes more efficient as $w \to -1$. By analysing the evolution of the fluid perturbations, we will demonstrate how this mechanism directly causes the observed shift in the suppression scale and explore its consequences for the growth of structure.

\subsection{Scale-dependent velocity-locking mechanism}

\begin{figure}[tbp]
\centering
\includegraphics[width=0.45 \textwidth]{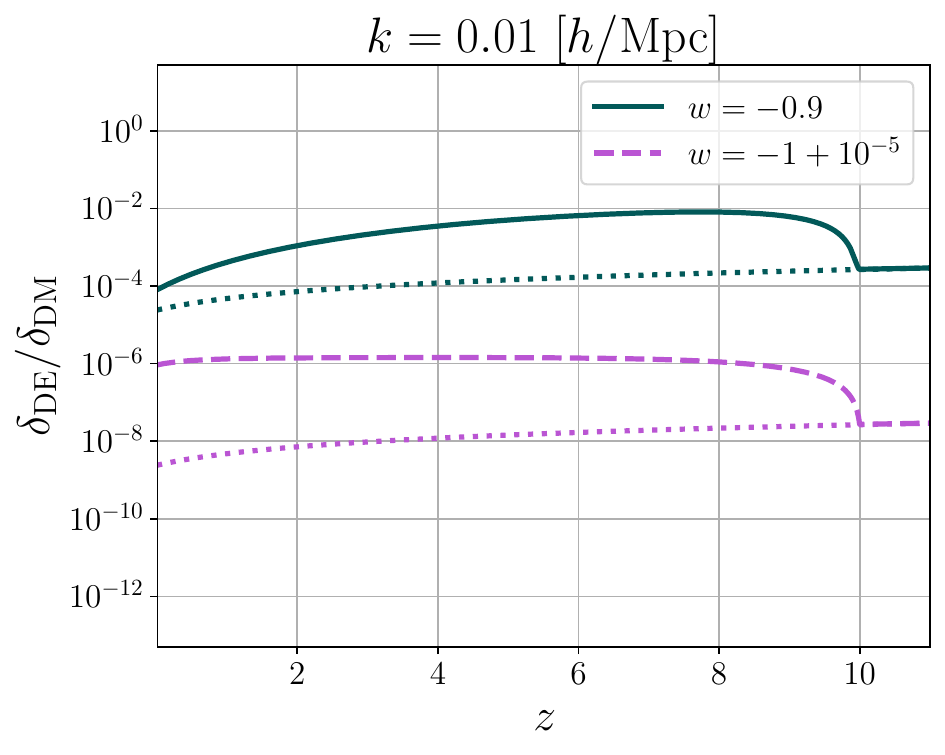}
\includegraphics[width=0.45 \textwidth]{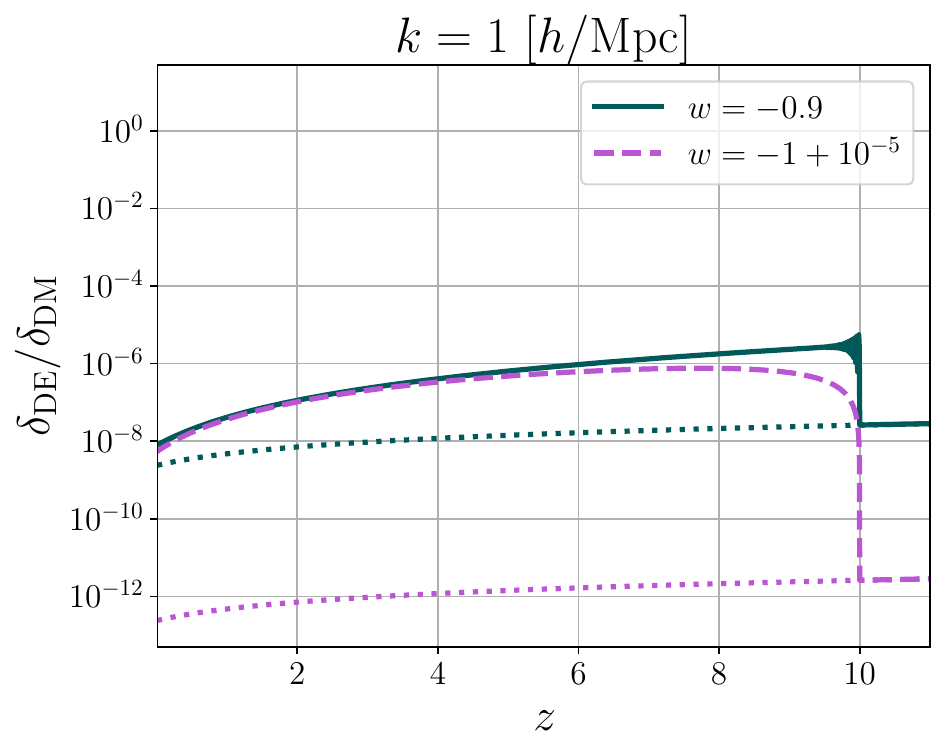}
\hfill
\caption{\label{fig:delta_de_over_delta_dm} 
Ratio of DE to DM fluid densities, $\delta_{\mathrm{DE}} / \delta_{\mathrm{DM}}$, as a function of redshift computed using the dark scattering model when $A = 15$ [b/GeV]. We choose to switch on the interaction at redshift $z=10$ and show the continuation of the non-interacting models at lower redshift (dotted lines). \textbf{Left:} $k = 0.01\,{\left[h/\mathrm{Mpc}\right]}$. \textbf{Right:} $k = 1\,{\left[h/\mathrm{Mpc}\right]}$. Solid lines show $w = -0.9$; dashed lines show $w = -1+10^{-5}$.}
\end{figure}

To understand the origin of the $w$-dependent suppression scale, we begin by examining the DE fluid equations in the $w \rightarrow -1$ limit. The first key insight comes from the DE continuity equation (Equation \ref{eq:delta_de}). As $w$ approaches $-1$, any term explicitly proportional to $(1+w)$ becomes negligible. Consequently, the equation reduces to $\delta_{\mathrm{DE}}' + 3(c_s^2 - w)\mathcal{H}\delta_{\mathrm{DE}} \approx 0$. This simplified form reveals that, in the absence of a significant source term from the velocity divergence, any existing DE density perturbation is subject to rapid decay due to Hubble friction. This smooths the DE fluid, rendering its density contrast, $\delta_{\mathrm{DE}}$, vanishingly small compared to when $w \neq -1$, as illustrated for different scales by the dotted lines in Figure \ref{fig:delta_de_over_delta_dm}. The sharp feature visible at $z = 10$ in the solid and dashed lines of the figure marks the initialisation of the interaction, which interrupts this decay. Tests with earlier initialisation redshifts, such as $z=30$, confirm that structure growth is negligibly impacted at high redshift because the dark energy density is subdominant to dark matter at these times.

\begin{figure}[tbp]
\centering
\includegraphics[width=0.45 \textwidth]{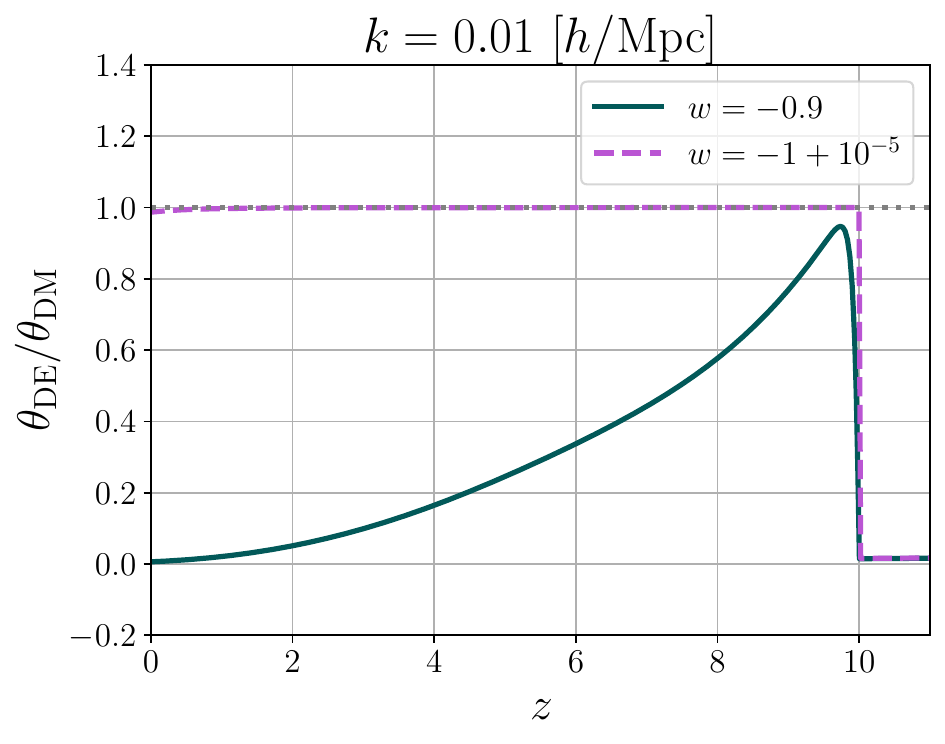}
\includegraphics[width=0.45 \textwidth]{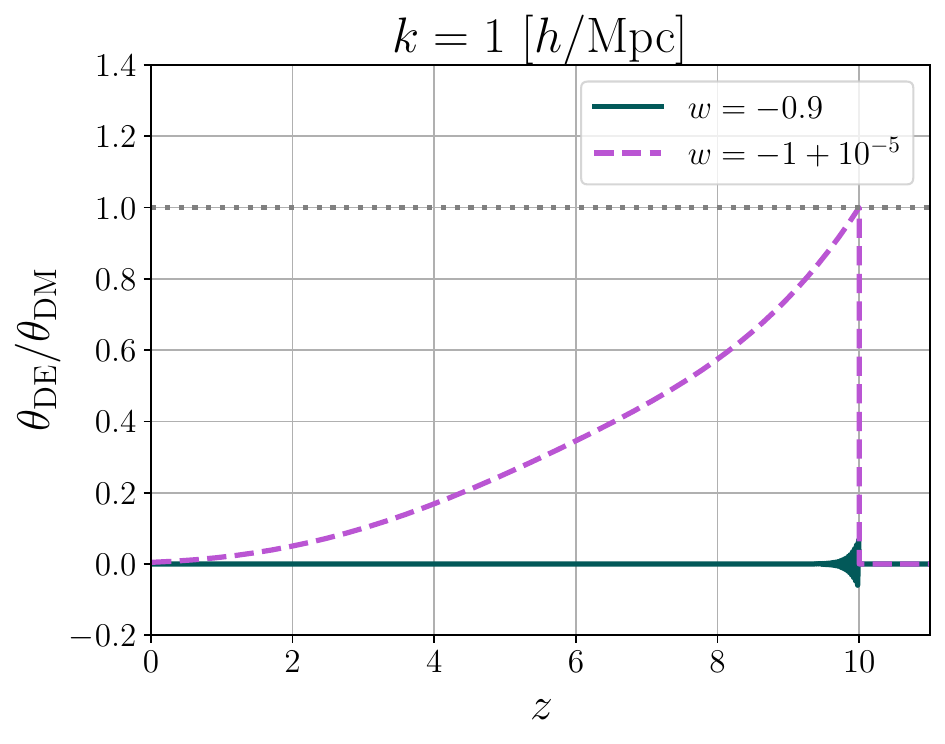}
\hfill
\caption{\label{fig:theta_de_over_theta_dm} 
Ratio of DE to DM fluid velocities, $\theta_{\mathrm{DE}} / \theta_{\mathrm{DM}}$, as a function of redshift computed using the dark scattering model when $A = 15$ [b/GeV]. We choose to switch on the interaction at redshift $z=10$. \textbf{Left:} $k = 0.01\,{\left[h/\mathrm{Mpc}\right]}$. \textbf{Right:} $k = 1\,{\left[h/\mathrm{Mpc}\right]}$. Solid lines show $w = -0.9$; dashed lines show $w = -1+10^{-5}$.
}
\end{figure}

A suppressed density contrast has a notable effect on the DE Euler equation (Equation \ref{eq:theta_de}). While the gravitational potential, $\psi$, sources perturbations in the DE fluid, the fluid's motion is primarily dictated by the competition between its internal pressure gradient, proportional to $c_s^2 k^2 \delta_{\mathrm{DE}}/(1+w)$, and the external drag from the interaction, proportional to $A(\theta_{\mathrm{DM}} - \theta_{\mathrm{DE}})/(1+w)$. Notably, while the coupling strength $A$ determines the efficiency of this drag and the resulting magnitude of the power spectrum suppression, it does not significantly alter the scale at which the unlocking occurs. Under typical conditions, where $w$ is not extremely close to $-1$, this competition is inherently scale-dependent. On large scales (low $k$), the pressure term is weak, allowing the interaction to couple the fluids, while on smaller scales the $k^2$ factor amplifies the pressure gradient, allowing it to overcome the drag and decouple the DE's motion. However, in the $w \rightarrow -1$ limit, the vanishingly small $\delta_{\mathrm{DE}}$ fundamentally weakens the pressure term's influence, even at high $k$. As a result, the powerful interaction drag dominates across a much broader range of scales than it otherwise would. This reinforcement of the velocity locking causes it to persist to much smaller physical scales (higher $k$), illustrated in Figure \ref{fig:theta_de_over_theta_dm}, fundamentally altering the scale-dependence of the interaction's impact.

A comparison of Figures \ref{fig:delta_de_over_delta_dm} and \ref{fig:theta_de_over_theta_dm} demonstrates how the underlying velocity dynamics dictate the density evolution. In Figure \ref{fig:delta_de_over_delta_dm}, the $w=-0.9$ model (solid lines) exhibits a consistent trajectory across both scales: the density ratio rises as the interaction pulls on the DE fluid to track the DM fluid, then decays as the fluids decouple. In contrast, the $w \approx -1$ model (dashed lines) displays a distinct scale dependence driven by the velocity locking visible in Figure \ref{fig:theta_de_over_theta_dm}. At large scales ($k=0.01\,\left[h/\mathrm{Mpc}\right]$), the drag force maintains a permanent velocity lock, which prevents decoupling and results in a constant density ratio. At small scales ($k=1\,\left[h/\mathrm{Mpc}\right]$), the density ratio initially surges to overlap with this large-scale amplitude, confirming that the physics becomes effectively scale-independent while the velocities are locked. However, as seen in Figure \ref{fig:theta_de_over_theta_dm}, this lock is temporary; the velocities eventually diverge as pressure support overcomes the drag. This breaking of the lock allows the density ratio to finally decay, mirroring the late-time fall-off of the $w=-0.9$ case.

This delay in the unlocking is the direct cause of the systematic shift seen in the power spectrum suppression. The drag on DM, governed by the interaction term in the DM Euler equation (Equation \ref{eq:theta_dm}), is only effective when the fluid velocities decouple ($\theta_{\mathrm{DE}} \neq \theta_{\mathrm{DM}}$). Consequently, this suppression effect is delayed until the DE pressure gradient is strong enough to finally break the velocity lock. With the magnitude of the DE density contrast, $|\delta_{\mathrm{DE}}|\propto(1+w)$, being so heavily suppressed in the $w \rightarrow -1$ limit, the pressure term requires a significantly larger $k$ for the $k^2$ factor to compensate and allow the pressure term to overpower the interaction term. This gives rise to a $w$-dependent unlocking scale. Equating the pressure and interaction terms shows this scale behaves as $k_\mathrm{unlock}^2 \propto \frac{\Gamma (\theta_\mathrm{DM}-\theta_\mathrm{DE})}{c_s^2(1+w)}$. While $\Gamma$ determines the interaction drag, the unlocking scale remains independent of the coupling strength because the velocity difference $(\theta_\mathrm{DM}-\theta_\mathrm{DE})$ adjusts inversely to the coupling strength in the locked regime. This leaves the total interaction drag magnitude effectively unchanged at these scales, leading to the fundamental relation $k_\mathrm{unlock} \propto 1/\sqrt{c_s^2(1+w)}$ and explaining why the onset of suppression moves to higher $k$ as $w \to -1$, as shown in Figures \ref{fig:pk_theta_de_off}, \ref{fig:pk} and \ref{fig:pk_theta_de_off_gamma}.

\subsection{Lowering the DE sound speed}

\begin{figure}[tbp]
\centering
\includegraphics[width=0.7 \textwidth]{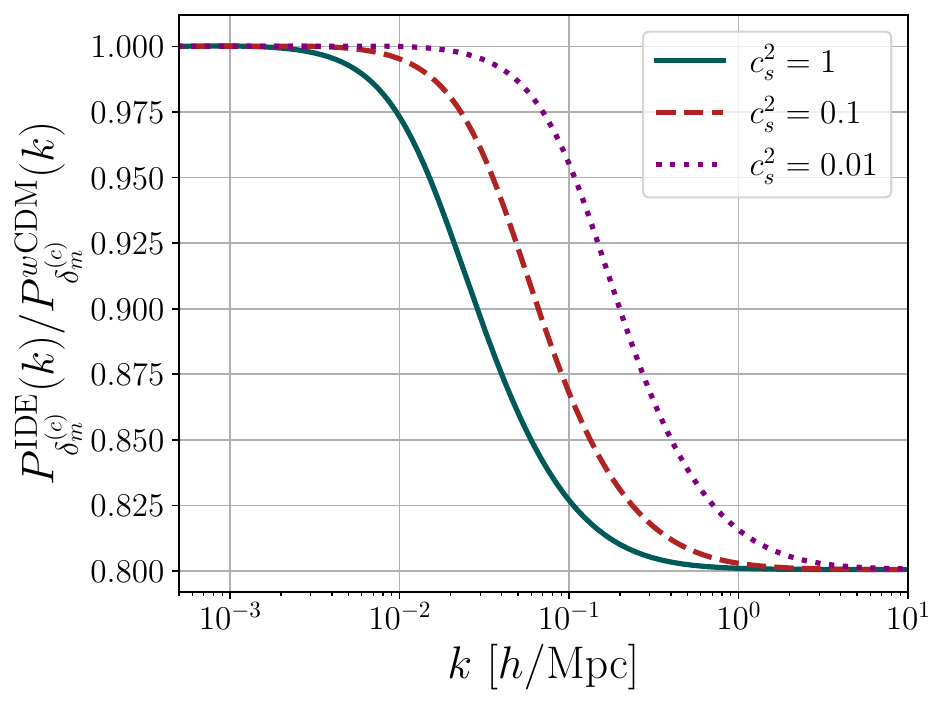}
\hfill
\caption{\label{fig:pk_cs2} 
Ratio of the linear matter power spectrum in the comoving-synchronous gauge with interaction to that of a non-interacting $w$CDM model, computed for $w = -1+10^{-3}$ and a variety of $c_s^2$ values at $z = 0$ using the dark scattering model when $A = 15$ [b/GeV].}
\end{figure}

The pressure gradient's effectiveness is also directly modulated by the DE sound speed, $c_s^2$. As is evident from the pressure term, a higher sound speed enhances the fluid's resistance to gravitational collapse, increasing the pressure support at all scales. This means that for a given value of $w$, a higher sound speed would allow the pressure gradient to overcome the interaction drag at a lower $k$, shifting the onset of suppression to larger scales. This dependence on sound speed is explicitly captured in the relation derived above ($k_\mathrm{unlock} \propto 1/\sqrt{c_s^2(1+w)}$), which shows that a smaller $c_s^2$ leads to a larger unlocking $k$ value. In this work, we have fixed the sound speed to $c_s^2 = 1$; however, we illustrate the effect of a range of sound speed values on the suppression onset scale in Figure \ref{fig:pk_cs2}.

\subsection[Impact of the \texorpdfstring{$\theta_{\mathrm{DE}}=0$}{thetaDE=0} approximation]{Impact of the \boldmath{$\theta_{\mathrm{DE}}=0$} approximation}
\label{sec:no_de}

\begin{figure}[tbp]
\centering
\includegraphics[width=0.45 \textwidth]{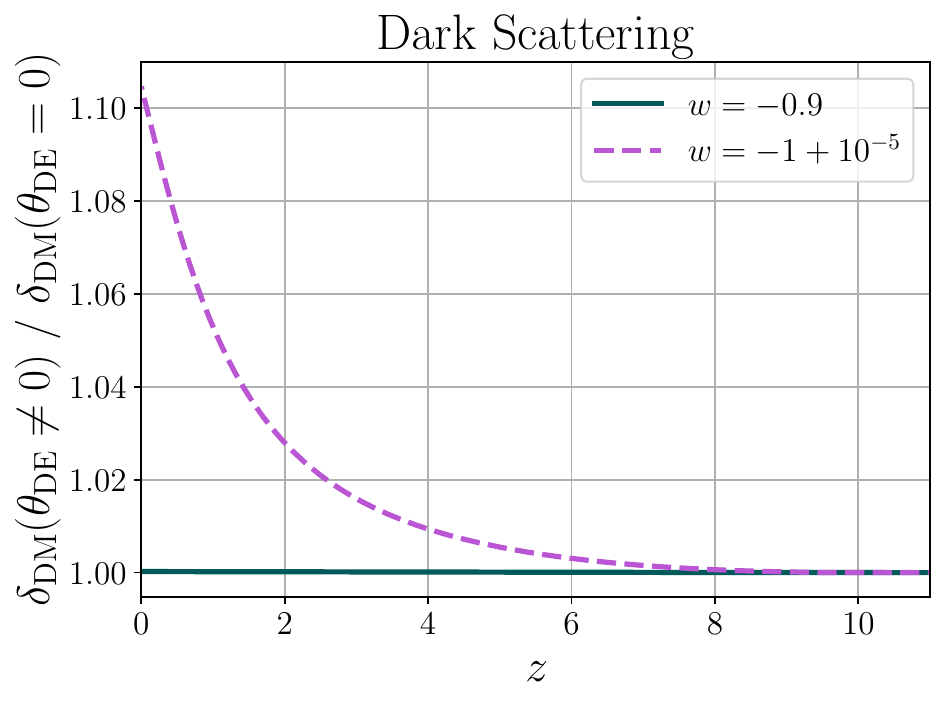}
\includegraphics[width=0.45 \textwidth]{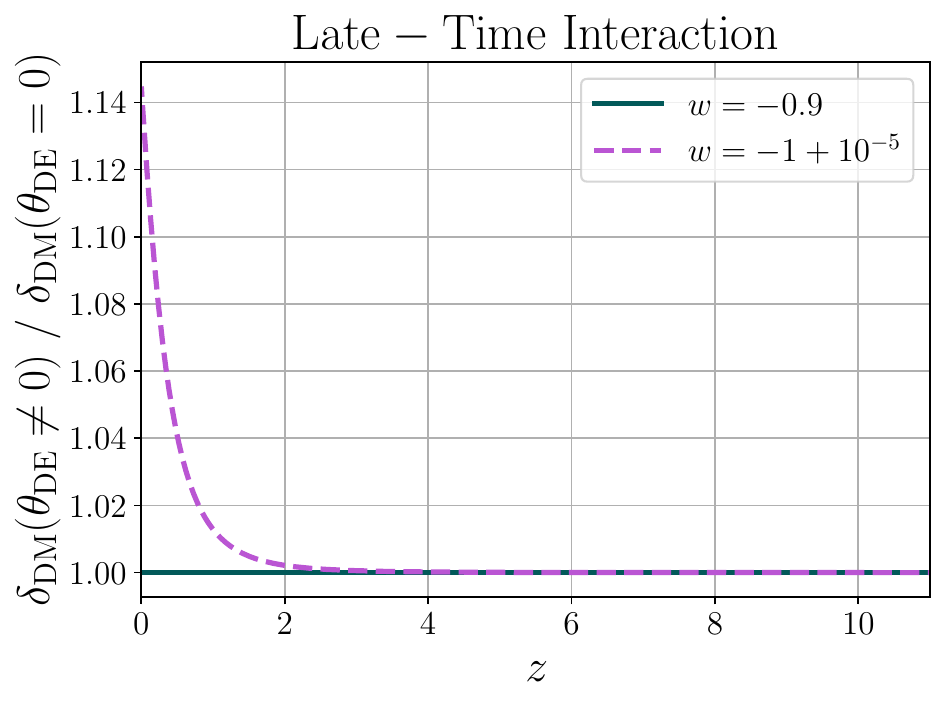}
\hfill
\caption{\label{fig:theta_on_off_deltas}
Ratio of $\delta_{\mathrm{DM}}$ computed with $\delta_{\mathrm{DE}}\ne0$ and $\theta_{\mathrm{DE}}\ne0$ to $\delta_{\mathrm{DM}}$ computed with $\delta_{\mathrm{DE}} = \theta_{\mathrm{DE}} = 0$, shown as a function of redshift. Results are shown for $k = 0.1\,{\left[h/\mathrm{Mpc}\right]}$ when $w = -0.9$ (solid) and $w = -1+10^{-5}$ (dashed). \textbf{Left:} The dark scattering model when $A = 15$ [b/GeV]. \textbf{Right:} The {late-time interaction} model when $\Gamma_{\mathrm{LT}} = 2{H_0}^3$.}
\end{figure}

A common approximation made in the literature is to treat DE fluctuations as negligible ($\delta_{\mathrm{DE}} = \theta_{\mathrm{DE}} = 0$) below the horizon scale, given the assumed sound speed of $c_s=1$ \cite{Baldi:2014ica,Baldi:2016zom,Bose:2017jjx,Carrilho:2021hly,Tsedrik:2022cri,Carrilho:2022mon,BeltranJimenez:2022irm}. While this can be a valid simplification in some contexts, it has a notable effect on the interaction in the $w \to -1$ limit. The right-hand panels of Figures \ref{fig:pk_theta_de_off} and \ref{fig:pk_theta_de_off_gamma} clearly show that when $\delta_{\mathrm{DE}}$ and $\theta_{\mathrm{DE}}$ are set to zero, the $w$-dependent shift of the unlocking scale is completely erased and the power spectrum suppression becomes scale-independent. Figure \ref{fig:theta_on_off_deltas} shows the ratio of the DM density contrast, when $\delta_{\mathrm{DE}}$ and $\theta_{\mathrm{DE}}$ are allowed to evolve, to the DM density contrast when $\delta_{\mathrm{DE}} = \theta_{\mathrm{DE}} = 0$. For $w = -0.9$, the ratio is unity, showing the approximation is highly accurate. For $w \approx -1$, however, the approximation introduces a large, redshift-dependent error. This error grows at late times and can exceed $10\%$ for both the dark scattering and late-time interaction models, demonstrating that the approximation fails significantly in this limit.

The reason for this deviation is that the approximation incorrectly assumes DE to be dynamically inert. As we have shown, the unlocking mechanism is governed by the dynamics of the DE velocity, $\theta_{\mathrm{DE}}$, as it competes with the interaction drag. Setting $\theta_{\mathrm{DE}}=0$ artificially maximises the velocity difference between the fluids ($\theta_{\mathrm{DM}} - \theta_{\mathrm{DE}} \approx \theta_{\mathrm{DM}}$), forcing the drag term to be fully active across all scales. This is physically incorrect and prevents the velocity-locking mechanism from emerging. As a consequence, using the $\delta_{\mathrm{DE}} = \theta_{\mathrm{DE}} = 0$ approximation for models where $w \approx -1$ leads to incorrect predictions about the scale-dependence of the interaction. This, in turn, could lead to a significant overestimation of the constraining power of observational data on the coupling parameters.

\section{Discussion and conclusion}

In this work, we have investigated the dynamics of pure momentum exchange models in the theoretically important and observationally relevant limit where the dark energy equation of state, $w$, approaches that of a cosmological constant. We have demonstrated that a generic feature of these models is a velocity-locking mechanism that becomes increasingly efficient as $w \to -1$. This mechanism arises from the interplay between the dark energy continuity and Euler equations. As $w$ nears $-1$, dark energy density perturbations are strongly suppressed, weakening the pressure gradient needed to decouple the dark fluids on small scales. As a result, the interaction's drag force dominates, locking the dark matter and dark energy velocities together over a wider range of scales. This systematically shifts the onset of matter power spectrum suppression to smaller scales (higher $k$), with the unlocking scale behaving as $k_{\mathrm{unlock}} \propto 1/\sqrt{1+w}$. In this limit, it is also demonstrated that a lower dark energy sound speed, $c_s^2$, further enhances this effect through the relation $k_{\mathrm{unlock}} \propto 1/\sqrt{c_s^2(1+w)}$. We have also shown that the common approximation of neglecting dark energy perturbations ($\delta_{\mathrm{DE}} = \theta_{\mathrm{DE}}=0$) fundamentally fails in this limit, as it artificially removes this physical mechanism and leads to incorrect predictions for the interaction's scale-dependence.

Our theoretical framework provides a direct physical explanation for recent observational results. Analyses by Laguë et al. \cite{Lague:2024sox} and the ACT collaboration \cite{ACT:2025tim} revealed that the constraining power on the interaction strength weakens significantly as $w$ approaches $-1$, a behaviour they attributed to a parameter degeneracy. The $w$-dependent velocity-locking mechanism identified in this work provides a physical explanation for these results, as it diminishes the model's observational signature in this regime. This underscores the importance of accurately modelling dark energy perturbations. As we have shown, the common approximation where $\delta_{\mathrm{DE}} = \theta_{\mathrm{DE}} = 0$ would erroneously erase this physical effect and lead to a significant overestimation of a model's constraining power.

The shift in power spectrum suppression to smaller scales as $w \rightarrow -1$ in the discussed fluid models also has an interesting phenomenological counterpart in Type 3 scalar field momentum interaction models \cite{Pourtsidou:2016ico,Chamings:2019kcl}. In these models, a similar effect is observed when the coupling parameter has a large negative value. As the magnitude of the coupling increases, the equation of state $w_\phi \rightarrow -1$. It has been shown that large negative values of the coupling parameter also result in an enhancement of the linear matter power spectrum relative to uncoupled quintessence for a range of $k$ values. The scale at which the power spectrum enhancement transitions to suppression shifts to higher $k$ as the magnitude of the coupling parameter increases. While there is not a direct correspondence between Type 3 and elastic scattering models, such phenomenological models can reproduce the characteristic drag-like behaviour of the scalar field models \cite{Skordis_2015,Baldi:2016zom}. Consequently, further investigation into these shared observational signatures will be required to determine the extent to which the phenomenological fluid approach captures the dynamics of Type 3 scalar field interactions.

Looking forward, beyond the comparison with Type 3 scalar field models, our findings open several other possibilities for further exploration into momentum-exchange dark sector interactions in the $w \to -1$ limit. This is increasingly important for upcoming Stage IV surveys, as current measurements indicate a potential phantom crossing \cite{DESI:2025zgx}, which could significantly affect the growth rate of structure in such interaction models. Crucially, while our analysis has focused on the linear regime, characterising how the velocity-locking mechanism shapes non-linear structure formation will be essential for translating this theoretical framework into concrete predictions testable by the next generation of surveys.

\acknowledgments

NC is supported by the UK Science and Technology Facilities Council (STFC) grant number ST/X508688/1 and funding from the University of Portsmouth. MB, RC and KK are supported by STFC grant number ST/W001225/1. KK is also supported by STFC grant number ST/B001175/1.

We acknowledge the use of the public code implementation for the late-time interaction model, described in \cite{Lague:2024sox} (\url{https://github.com/fmccarthy/Class_DMDE}). We thank the anonymous referee for the helpful suggestions.

For the purpose of open access, we have applied a Creative
Commons Attribution (CC BY) licence to any Author Accepted
Manuscript version arising. Supporting research data are available on reasonable request from the authors.

\bibliographystyle{JHEP}
\bibliography{references.bib}

\end{document}